# Temperature compensation via cooperative stability in protein degradation

Yuanyuan Peng[1]*, Yoshihiko Hasegawa[2], Nasimul Noman[3], Hitoshi Iba[2]

[1]Research Center for Advanced Science and Technology, The University of Tokyo, Tokyo, Japan

[2]Graduate School of Information Science and Technology, The University of Tokyo, Tokyo, Japan

[3]School of Electrical Engineering and Computer Science, The University of Newcastle, Callaghan NSW 2308, Australia

*Corresponding author

Email Addresses:

    YP: peng@iba.t.u-tokyo.ac.jp

    YH: hasegawa@iba.t.u-tokyo.ac.jp

    NN: nasimul.noman@newcastle.edu.au

    HI: iba@iba.t.u-tokyo.ac.jp

**Abstract**

Temperature compensation is a notable property of circadian oscillators that indicates the insensitivity of the oscillator system's period to temperature changes; the underlying mechanism, however, is still unclear. We investigated the influence of protein dimerization and cooperative stability in protein degradation on the temperature compensation ability of two oscillators. Here, cooperative stability means that high-order oligomers are more stable than their monomeric counterparts. The period of an oscillator is affected by the parameters of the dynamic system, which in turn are influenced by temperature. We adopted the Repressilator and the Atkinson oscillator to analyze the temperature sensitivity of their periods. Phase sensitivity analysis was employed to evaluate the period variations of different models induced by perturbations to the parameters. Furthermore, we used experimental data provided by other studies to determine the reasonable range of parameter temperature sensitivity. We then applied the linear programming method to the oscillatory systems to analyze the effects of protein dimerization and cooperative stability on the temperature sensitivity of their periods, which reflects the ability of temperature compensation in circadian rhythms. Our study explains the temperature compensation mechanism for circadian clocks. Compared with the no-dimer mathematical model and linear model for protein degradation, our theoretical results show that the nonlinear protein degradation caused by cooperative stability is more beneficial for realizing temperature compensation of the circadian clock.

**Keywords:** Genetic oscillator, Cooperative stability, Phase sensitivity analysis, Temperature compensation, Linear programming

## 1. Introduction

Circadian clocks keep their periods almost unchanged when the temperature varies. This robustness against temperature variation, a famous mechanism in circadian clocks, is known as temperature compensation [1-9]. Although the period of the circadian clock is insensitive to thermal variations, the rates of reactions such as synthesis and degradation of mRNA and proteins are highly temperature dependent [1, 10]. However, the mechanisms by which the circadian rhythms compromise several reaction steps to realize temperature compensation are still unknown. Thermal variation results in differences in the reaction parameters of the dynamical systems, which in turn should change the period of the oscillators. The prevalence of the cooperative processes in nature inspired us to investigate the relationship between various levels of cooperation and the temperature sensitivity of the oscillators' period. Most previous research has been focused on cooperation at the stage of transcription [10-13], which is quite limited, and it was often difficult for the circuits to perform oscillation in the range of physiological parameter values because of insufficient cooperation. Hence, it is necessary to harness cooperativity during other processes of gene expression, such as translation and protein degradation [14].

Buchler *et al.* studied cooperation in protein degradation, and pointed out that nonlinear protein degradation achieved by cooperative stability can widen the oscillation parameter space [15]. Here, cooperative stability means that dimers or high-order oligomers are more stable to proteolysis than monomers. Hong and Tyson proposed a molecular mechanism for temperature compensation based on the opposing effects of temperature on the rate of nuclear import of period (PER) protein and the association rate of PER monomers [16]. But they did not consider the physiological range of the temperature sensitivity of the parameters and the effects of temperature on the other reaction parameters, such as the synthesis and degradation rates of mRNA, monomers, and dimers. We analyzed the influence of protein dimerization and cooperative stability on the temperature compensation ability of circadian clocks taking these problems into account. Biological oscillators can be classified into two types: (1) smooth oscillators containing only negative feedback loops; and (2) relaxation oscillators including both positive and negative feedback loops [17]. Circadian clocks, as special biological oscillators, belong to one of these two types, and have the basic characteristics of these oscillators. Thus, we can analyze the temperature compensation ability of the circadian rhythms by considering smooth and relaxation oscillators instead of circadian clocks. We used the Repressilator and the Atkinson oscillator [18] to analyze period robustness against temperature changes. The Repressilator is a smooth oscillator, while the Atkinson oscillator is a relaxation oscillator. Despite their simplicity in topologies, these oscillators can exhibit rich dynamical behaviors and have many properties in common with genetic oscillators [11, 18-22]. Therefore, when the environmental temperature varies, the changes in the periods of the two oscillators with different mechanisms can uncover the influence on the temperature compensation ability. We analyzed the temperature sensitivity of the period for three cases using the linear programming method. Specifically, we used the mathematical models without protein dimerization, and linear and nonlinear protein degradation models for both the Repressilator and the Atkinson oscillator. The period's temperature sensitivity was adopted to classify whether the temperature compensation ability was strong or weak [1, 23].

The temperature sensitivity of the period depends on two factors: the period sensitivity and the temperature sensitivity of the parameters. Phase sensitivity analysis can measure the deviations in period induced by perturbations to the reaction parameters of the systems [24-26], which are the parameters' period sensitivity needed for the calculation of the temperature sensitivity of the period. The values of the parametric temperature sensitivities have a special range according to recently provided experimental data [27]. Thus, we can obtain the best result for the minimum temperature sensitivity of the period of the oscillators by using linear programming. Our main findings are that protein dimerization and cooperative stability can improve the temperature compensation ability of the oscillators. When the temperature sensitivity of the period is higher in the oscillators, temperature compensation ability is weaker; conversely, a lower value implies a stronger temperature compensation ability. To our knowledge, this is the first report of using linear programming to evaluate the temperature compensation ability of biochemical oscillators.

## 2. Mathematical Models of Genetic Oscillators

### 2.1 Protein Dimerization and Cooperative Stability for the Oscillators

Protein degradation substantially affects the functional properties of genetic circuits, and ample experimental evidence suggests that many proteins are functional in the form of dimers or even higher order oligomers [28, 29]. The stability of oligomers to proteolysis is higher than that of monomers [30, 31], and this enhanced stability is referred to as cooperative stability [15]. We studied the influence of protein dimerization and cooperative stability on the properties of two kinds of genetic oscillators: the Repressilator and the Atkinson oscillator. Although these two oscillators have been experimentally implemented in *Escherichia coli*, they exhibit oscillatory dynamics via different mechanisms. The three repressors of the Repressilator are connected in a ring topology, and the expression of each gene is inhibited by its downstream partner, forming a negative feedback loop. The Atkinson oscillator organizes repression and activation in the gene network to regulate the oscillation function. According to experimental results, the oscillation of the Repressilator disappears after a short time [19], whereas the Atkinson oscillator can maintain damped oscillation for a relatively long time [18]. The Repressilator and the Atkinson oscillator represent the smooth oscillator and relaxation oscillator according to their topologies, respectively. We considered the generic effects of protein dimerization and cooperative stability on the characteristics of these two types of oscillators.

Figure 1B illustrates the gene expression, including the effect of cooperative stability, through which we can describe the processes of transcription, translation, dimerization, and degradation along with several indispensable kinetic parameters [15]. We also used a mathematical model with only monomers (Fig. 1A) to determine the effect of protein dimerization. The parameters illustrated in Figs. 1A and B are defined as follows: $\alpha$ represents the transcriptional rate at full activation; $v$ is the translation rate for proteins; and $\lambda_{p1}$, $\lambda_{p2}$, and $\lambda_m$ indicate the degradation rates for monomer ($p_1$), dimer ($p_2$), and mRNA ($m$), respectively. The mRNA synthesis rate in bacteria is often repressed or activated by transcription factors (TF) that usually function in the form of monomers or homodimers; hence, the relative activation and repression functions of promoter activities in our study are represented by the Hill functions $g_a$(TF) and $g_r$(TF) (TF: $p_1$ or $p_2$), which are functions of the monomer ($p_1$) or dimer ($p_2$) concentrations. Figures 1C and D illustrate the gene networks of the Repressilator and the Atkinson oscillator, respectively.

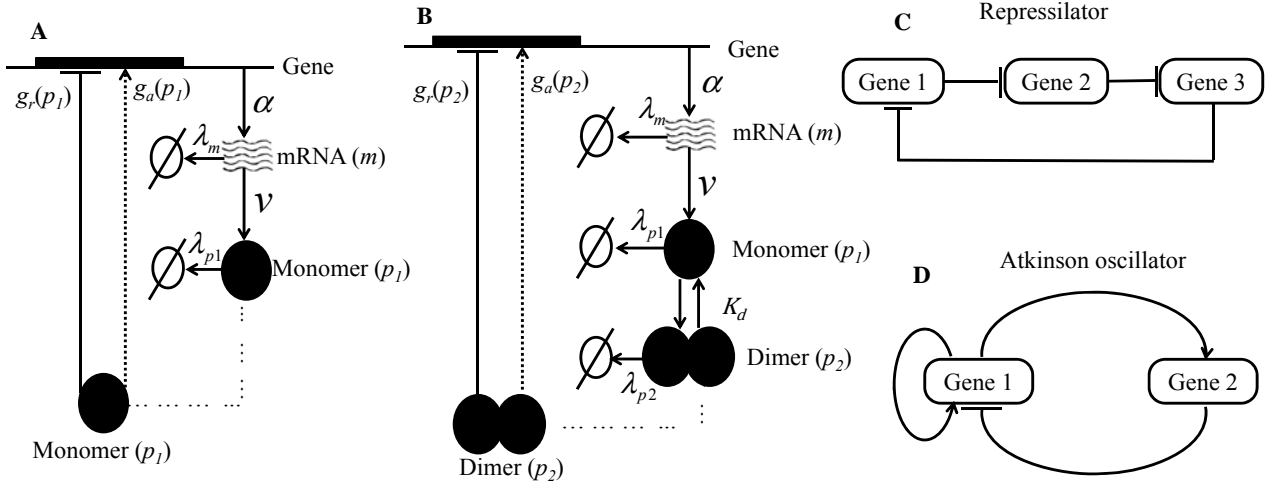

**Fig. 1.** Gene expression and gene networks of the Repressilator and the Atkinson oscillator. (A) Schematic diagram of the gene expression without dimerization. (B) Schematic diagram of the regulation from transcription to dimerization via translation. The promoter in the schematic diagrams has two possible activities: the solid blunt line indicates repression with transcription rate $\alpha \cdot g_r(p_1)$ or $\alpha \cdot g_r(p_2)$, and the dashed line with arrows indicates active regulation of the promoter for transcription with a rate $\alpha \cdot g_a(p_1)$ or $\alpha \cdot g_a(p_2)$. The functions $g_r$(TF) and $g_a$(TF) (TF: $p_1$ or $p_2$) are the Hill functions. The mRNA ($m$) decays at rate $\lambda_m$, and the protein monomers ($p_1$) are synthesized at rate $v$ and decay at a rate of $\lambda_{p1}$. The concentration of dimers ($p_2$) depends on the equilibrium dissociation constant $K_d$ and decay rate $\lambda_{p2}$. (C) Gene network for the Repressilator. The expression of one gene is repressed by another gene in the same network. (D) Gene circuit for the Atkinson oscillator. Protein encoded by gene 1 plays an active role in promoting the transcription of gene 1 and gene 2, and the protein encoded by gene 2 inhibits the expression of gene 1.

## 2.2 Models of the Repressilator and the Atkinson Oscillator

According to the previous description of gene expression, we can model the two genetic oscillators using ordinary differential equations (ODEs) describing the net change in the mRNA and protein concentration caused by transcription, translation, and degradation.

The ODEs for the Repressilator with only monomers as transcription factors in the Hill functions are

$$\frac{dm^{(i)}}{dt} = \alpha g_r\big(p^{(i-1)}\big) - \lambda_m m^{(i)},$$

$$\frac{dp^{(i)}}{dt} = vm^{(i)} - \lambda_{p1} p^{(i)}, \tag{1}$$

$$i = 1, 2, 3, \ \ p^0 = p^3.$$

The ODEs for the Repressilator with cooperative stability for every protein are

$$\frac{dm^{(i)}}{dt} = \alpha g_r\left(p_2^{(i-1)}\right) - \lambda_m m^{(i)},$$

$$\frac{dp^{(i)}}{dt} = vm^{(i)} - \left(\lambda_{p1} p_1^{(i)} + 2\lambda_{p2} p_2^{(i)}\right), \quad (2)$$

$$i = 1, 2, 3, \quad p_2^0 = p_2^3.$$

The ODEs for the Atkinson oscillator with only monomers as transcription factors in the Hill functions are

$$\frac{dm^{(1)}}{dt} = \alpha g_r(p^{(2)}) g_a(p^{(1)}) - \lambda_m m^{(1)},$$

$$\frac{dm^{(2)}}{dt} = \alpha g_a(p^{(1)}) - \lambda_m m^{(2)},$$

$$\frac{dp^{(i)}}{dt} = vm^{(i)} - \lambda_{p1} p^{(i)}, \quad (3)$$

$$i = 1, 2.$$

The ODEs for the Atkinson oscillator with cooperative stability for each protein are

$$\frac{dm^{(1)}}{dt} = \alpha g_r(p_2^{(2)}) g_a(p_2^{(1)}) - \lambda_m m^{(1)},$$

$$\frac{dm^{(2)}}{dt} = \alpha g_a(p_2^{(1)}) - \lambda_m m^{(2)},$$

$$\frac{dp^{(i)}}{dt} = vm^{(i)} - \left(\lambda_{p1} p_1^{(i)} + 2\lambda_{p2} p_2^{(i)}\right), \quad (4)$$

$$i = 1, 2.$$

In each case, $m^{(i)}$ and $p^{(i)}$ represent the concentrations of the $i$th mRNA and total protein, respectively. The total protein concentration includes the concentration of the dimer ($p_2$) and monomer ($p_1$), that is, $p^{(i)} = p_1^{(i)} + 2p_2^{(i)}$, and for Eqs. (1) and (3), $p_2^{(i)} = 0$. According to Eqs. (2) and (4), when the degradation rate of the monomeric protein $\lambda_{p1}$ is equal to the decay rate of the dimeric protein $\lambda_{p2}$, that is, $\lambda_{p1} = \lambda_{p2}$, the degradation rate for the total protein ($p$) is linear, while $\lambda_{p1} > \lambda_{p2}$ corresponds to nonlinear degradation of total protein concentration. The nonlinear case leads to cooperative stability in protein degradation. According to Figs. 1A and B, the synthesis rate of monomers at full activation can be calculated by $\gamma = \alpha v / \lambda_m$ for both the Repressilator and the Atkinson oscillator. When the concentrations of the monomers and dimers reach equilibrium (Fig. 1B) rapidly, the relationship between them can be expressed as $p_2 = p_1^2 / K_d$, where $K_d$ is the equilibrium dissociation constant; see reference [15] and its supplementary material for more information about the formation and degradation of the monomer and dimer. In Eqs. (1)–(4), the positive feedback $g_a(\text{TF})$ and negative feedback $g_r(\text{TF})$ of the promoter are represented by Hill functions of monomeric ($p_1$) or homodimeric ($p_2$) concentration [15, 32-34]:

$$g_a(\text{TF}) = \frac{l^{-1} + \left(\frac{\text{TF}}{k}\right)^n}{1 + \left(\frac{\text{TF}}{k}\right)^n},$$

$$g_r(\text{TF}) = \frac{1 + \left(\frac{\text{TF}}{k}\right)^n/l}{1 + \left(\frac{\text{TF}}{k}\right)^n},$$

where $l$ denotes an $l$-fold change from the basal to the maximal value of the function, $n$ indicates the degree of cooperativity for the Hill function, and the concentration of protein separating the transition region from the saturation level is expressed as $k$. Apart from the two cases for the two oscillators—dimerization and no dimerization—expressed by Eqs. (1)–(4), we also consider two other situations in Appendix A to verify the effects of dimerization and cooperative stability: we vary the number of proteins having cooperative stability for the Repressilator; and only $p^1$ or $p^2$ in the Atkinson oscillator can show cooperative stability.

The values of the parameters in this study are physiologically realizable in bacteria, and thus reflect real biological situations. Table 1 summarizes the description and feasible values of the parameters. For more details on the parameters, refer to [15] and the references therein.

**Table 1.** Key parameter values for the Repressilator and the Atkinson oscillator. All the values in the table fall in the physiological ranges for bacteria.

| Parameters | Repressilator | Atkinson |
|---|---|---|
| $l$ | 1000 | 100 |
| $k$ | 3 nM | 6 nM |
| $n$ | 2 | 2 |
| $\gamma$ | 50–200 nM/min | 50–200 nM/min |
| $\lambda_{p1}$ | 0.2 min$^{-1}$ | 0.2 min$^{-1}$ |
| $\lambda_{p2}$ | Linear: (0.2); nonlinear: (0.04/0.02) min$^{-1}$ | Linear: (0.2); nonlinear: (0.04/0.02) min$^{-1}$ |
| $K_d$ | 10 nM | 10 nM |

In the following section, we analyze the influence of protein dimerization and nonlinear protein degradation caused by cooperative stability on the properties of the Repressilator and the Atkinson oscillator. The analysis focuses on mainly the changes of the oscillators' periods caused by thermal variation in different mathematical models.

## 3. Phase Sensitivity Analysis for the Oscillators

The phase reduction method can reduce the high-dimensional ODEs expressing the dynamical systems of the genetic oscillators to a single ODE, while retaining properties of the systems, such as the phase and period, [35, 36]. Through phase sensitivity analysis, we can understand how applying an infinitesimal perturbation to the parameters affects period of the oscillators.

The following ODE describes the dynamics of genetic oscillators:

$$\frac{d\boldsymbol{y}}{dt} = \boldsymbol{f}(\boldsymbol{y}(t), \boldsymbol{b}), \tag{5}$$

where $\boldsymbol{y} \in \boldsymbol{R}^N$ represents the vector of $N$ states of the system, for example, the concentration of protein and mRNA; $\boldsymbol{b} \in \boldsymbol{R}^M$ denotes the vector of the $M$ reaction rate parameters, such as the synthesis and degradation rates of mRNA and protein; $\boldsymbol{f}$ indicates the function of the parameters and states; and $t$ is time. The orbit of the oscillator is denoted as $\varsigma$, and the solution along the trajectory is represented by $\boldsymbol{y}^\varsigma(t)$, where $\boldsymbol{y}^\varsigma(t) = \boldsymbol{y}^\varsigma(t + \tau)$ always holds ($\tau$ is the period). We defined the phase of the point $\boldsymbol{y}^\varsigma$ on the trajectory of the oscillators in Appendix B in order to perform the phase sensitivity analysis.

An important concept for phase sensitivity analysis is the phase response curve (PRC). The set of phase shifts induced by small short-lived stimuli at different times (phases) of the orbit is the PRC, and the symbol $\boldsymbol{U}$ represents a vector of the PRCs caused by impulse perturbations to all the states of the oscillators. Although the PRC is the simplest phase analysis, it is necessary for studying more complex phase sensitivities. We can compute the PRCs $\boldsymbol{U}$ for the different models of the Repressilator and the Atkinson oscillator according to the theory provided in Appendix B, and the computation results of PRCs for the two oscillators provide a strong foundation for the following analysis of period sensitivity.

### 3.1 Parametric Sensitivity Analysis

The analysis of the phase responses caused by stimuli to the parameters is more difficult than the analysis of the PRCs ($\boldsymbol{U}$ expressed by Eq. (B3)) because the variations of the parameters lead to orbits different from the nominal one. Taylor *et al.* [24] measured the phase shifts according to the time difference between the perturbed and nominal limit cycles to reach the same isochron. The following equation provides the theoretical calculation method for PRCs caused by impulse perturbations to the parameters (pIPRC) $\boldsymbol{Z}$ (see the appendix of [24] for its derivation):

$$Z_j(\mathbf{y}^c(t)) = \sum_{i=1}^{N} U_i(\mathbf{y}^c(t)) \cdot \frac{\partial f_i}{\partial b_j}(\mathbf{y}^c(t)), \tag{6}$$

where $N$ is the number of equations describing the dynamical system, $\frac{\partial f_i}{\partial b_j}$ represents the partial differential with respect to the parameter $b_j (j = 1, 2, ..., M)$ of the function $\mathbf{f}$ in Eq. (5), and $Z_j$ and $U_i$ denote the pIPRC and PRC, respectively. The parametric phase sensitivity calculated by Eq. (6) reflects the cumulative phase difference between the perturbed and unperturbed states, that is, the PRC. Ample experimental data has confirmed that environmental temperature changes affect the mRNA synthesis rate $\alpha$ and degradation rate $\lambda_m$, the protein translation rate $v$, the monomer degradation rate $\lambda_{p1}$, and the dimer degradation rate $\lambda_{p2}$ [37]; hence, we only considered the pIPRC for the parameters mentioned previously.

## 3.2 Normalized Period Sensitivity

The pIPRC, caused by the pulse perturbations to the parameters of the systems, reflects the direction and value of the phase variation deviating from the nominal limit cycle. Period sensitivity provides a useful way to evaluate the extent of the change in the free-running period when the duration of the stimuli to the parameters lasts a long time. Period sensitivity is defined as the accumulation of numerous phase sensitivities caused by short-lived perturbations to the parameters during one period; we can analytically express the relationship between period sensitivity and pIPRC $Z_j (j = 1, 2, ..., M)$ as

$$\frac{\partial \tau}{\partial b_j} = -\int_{t_0}^{t_0+\tau} Z_j(\mathbf{y}^c(t)) dt, \tag{7}$$

where $\tau$ represents the period of the oscillator, and $t_0$ is any point on the trajectory. Eq. (7) tells us that the period sensitivity $\frac{\partial \tau}{\partial b_j}$ equals the area under the pIPRC $Z_j$ in one period, but the different parametric values and periods of the oscillators make it difficult to compare period sensitivities. To consider the robustness of the period based on identical standards, we used the normalized period sensitivity provided by

$$e_j = \frac{\partial \ln \tau}{\partial \ln b_j} = \frac{b_j}{\tau} \cdot \frac{\partial \tau}{\partial b_j}, \tag{8}$$

where the rate parameter $b_j (j = 1, 2, ..., M)$ and the period of the limit cycle $\tau$ are known values, and $\frac{\partial \tau}{\partial b_j}$ is the period sensitivity calculated in Eq. (7).

From the theory described by Eqs. (7)–(8), we can calculate the normalized period sensitivity to different parameters in the two oscillators for the linear and nonlinear protein degradation models and the no-dimer model. Figure 2 shows the computational results of the normalized period sensitivities for the models (Eqs. (1)–(4)) as a function of the protein synthesis rate. The normalized period sensitivity results for Eqs. (2) and (4) include linear ($\lambda_{p1} = \lambda_{p2}$) and nonlinear ($\lambda_{p1} = 5\lambda_{p2}$, $\lambda_{p1} = 10\lambda_{p2}$) degradation rates. The protein synthesis rate $\gamma$ at full activation is selected to be 50–200 nM/min for both the Repressilator and the Atkinson oscillator, over which the systems exhibit oscillations. To obtain the previously described range for the protein synthesis rate, defined by $\gamma = \alpha \cdot v/\lambda_m$, the values for $\alpha$, $v$, and $\lambda_m$ are set to 50–200 nM/min, 0.2 min$^{-1}$, and 0.2 min$^{-1}$ for the Repressilator, and 5–20 nM/min, 2 min$^{-1}$, and 0.2 min$^{-1}$ for the Atkinson oscillator, respectively. Table 1 lists the other parameters used for the computation. The normalized period sensitivity for the mathematical models expressed by Eqs. (1)–(4) in cases when the translation rate $v$ and the degradation rate $\lambda_m$ vary is given in the Supplemental Material. We also provide normalized period sensitivity data as a function of protein synthesis rate $\gamma$ for the models with only one or two proteins showing cooperative stability given by Eqs. (A1)–(A4) in the appendix.

Figure 2 shows normalized period sensitivities to the transcription rate $\alpha$ (A and F), translation rate $v$ (B and G), degradation rate of mRNA $\lambda_m$ (C and H), monomer decay rate $\lambda_{p1}$ (D and I), and dimer decay rate $\lambda_{p2}$ (E and J). The left figures represent the results for the Repressilator, and the right are for the Atkinson oscillator. The solid (blue), dashed (red), dotted (green), and dash-dot (black) lines represent the normalized period sensitivity of the linear protein degradation model, the nonlinear protein degradation model with $\lambda_{p1} = 5\lambda_{p2}$, the nonlinear protein degradation model with $\lambda_{p1} = 10\lambda_{p2}$, and the model with only monomers, respectively. The Repressilator's normalized period sensitivities for reaction rate $\alpha$ (Fig. 2A), $v$ (Fig. 2B), $\lambda_m$ (Fig. 2C), and $\lambda_{p2}$ (Fig. 2E) indicate that the influence of the perturbations on the period of the nonlinear protein degradation model is less than that on the period of the linear protein degradation model; however, the period of the nonlinear protein degradation model has almost the same response as that of the linear model when there are small variations to the monomer protein decay rate (Fig. 2D). From Figs. 2A, C, and E, we can conclude that the period of the oscillator with the relationship $\lambda_{p1} = 10\lambda_{p2}$ is more robust to the disruption of parameters $\alpha$, $\lambda_m$, and $\lambda_{p2}$ than that with the smaller nonlinearity in protein degradation ($\lambda_{p1} = 5\lambda_{p2}$). The degree of nonlinearity does not make much difference in the response to perturbations of parameters $v$ and $\lambda_{p1}$ (dashed (red) and dotted (green) lines) shown in Figs. 2B and D. If there are no dimers in the network topology of the Repressilator, the period sensitivity to $\lambda_{p1}$ is much greater than that of the linear and nonlinear protein degradation models (Fig. 2D). Figures 2A, B, and C show that the period sensitivity to $\alpha$, $v$, and $\lambda_m$ in the mathematical model with no dimers mostly falls between those of the linear and nonlinear protein degradation models.

The period sensitivity of the Atkinson oscillator is more complex. Figures 2H and J showing the normalized period sensitivity to $\lambda_m$ and $\lambda_{p2}$ indicate that higher robustness was exhibited for the nonlinear protein degradation model than for the linear protein decay model. Figures 2F and G show that when the protein synthesis rate $\gamma$ was slow, the period of the nonlinear Atkinson oscillator was more robust than that of the linear model in resisting the parametric perturbations to the transcription rate $\alpha$ and the protein translation rate $v$; however, as $\gamma$ increased, the nonlinear model lost its robustness. Large differences in the period changes caused by stimuli to $\lambda_{p1}$ in the Atkinson oscillator (Fig. 2I) were not observed between the linear and nonlinear models in most regions of the protein synthesis rate. When the protein synthesis rate was high, the nonlinear protein degradation model with $\lambda_{p1} = 10\lambda_{p2}$ tended to generate a larger period sensitivity to parameters $\alpha$, $v$, $\lambda_m$, and $\lambda_{p1}$ than the nonlinear model with $\lambda_{p1} = 5\lambda_{p2}$. When the protein synthesis rate was low, the nonlinearity did not cause large differences to the period sensitivity to these parameters (Figs. 2F–I). Figure 2J shows that the period of the model with a smaller dimeric protein degradation rate (dotted-green line) was more difficult to change than that of the model with the larger dimeric protein degradation rate (dashed-red line). The period sensitivities of the mathematical models without dimers (dash-dotted (black) lines in Figs. 2F–I) were much flatter in the physiological range of protein synthesis rate $\gamma$ than those of the linear and nonlinear protein degradation models. Compared to the linear and nonlinear protein degradation models, the model without protein dimerization exhibited a period that was more sensitive to $v$, $\lambda_m$, and $\lambda_{p1}$ (Figs. 2G, H, and I) in most situations. The perturbation to the transcription rate $\alpha$ in the no-dimer model affected the period less than in the linear and nonlinear protein degradation models (Fig. 2F). We also investigated the normalized period sensitivities of Eqs. (1)–(4) as a function of $v$ and $\lambda_m$, as shown in Figs. S1 and S2 in the supplemental material. The results are similar to those illustrated in Fig. 2. The normalized period sensitivities as a function of the protein synthesis rate for the mathematical models in Eqs. (A1)–(A4) are shown in Figs. S3–S8.

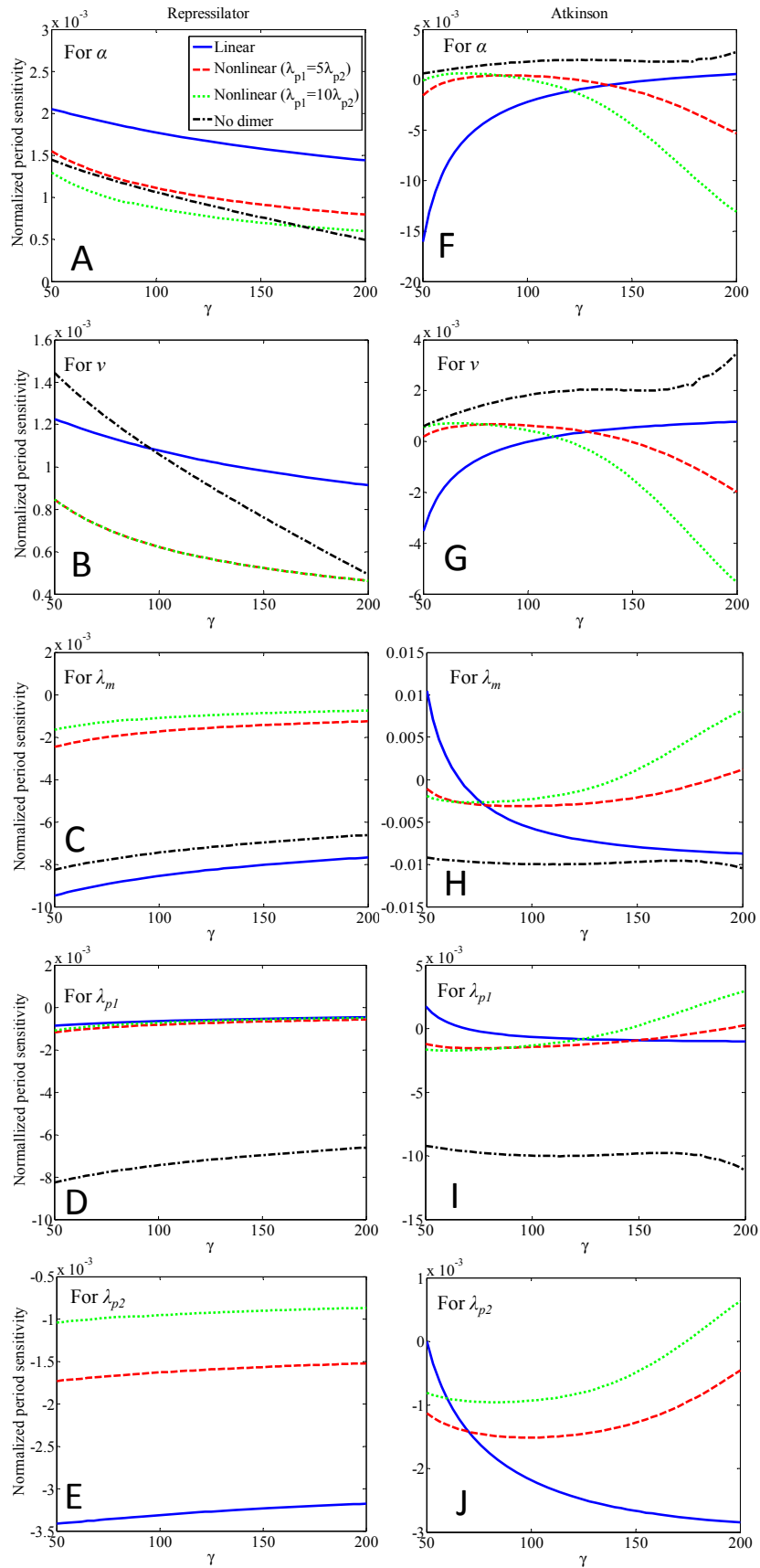

**Fig. 2.** Normalized period sensitivities $e_j$ to the reaction rates in the linear/nonlinear protein degradation (Eqs. (2) and (4)) and no–dimer (Eqs. (1) and (3)) models of the oscillators when $\alpha$ varies. Sensitivities to (A) and (F) the mRNA synthesis rate $\alpha$; (B) and (G) the protein monomer translation rate $v$; (C) and (H) the mRNA degradation rate $\lambda_m$; (D) and (I) the monomer protein degradation rate $\lambda_{p1}$; and (E) and (J) dimer protein degradation $\lambda_{p2}$. The solid (blue), dashed (red), dotted (green), and dash-dotted (black) lines represent the period sensitivity in four different cases, namely, the linear protein degradation ($\lambda_{p1} = \lambda_{p2}$), nonlinear protein degradation ($\lambda_{p1} = 5\lambda_{p2}$), nonlinear protein degradation ($\lambda_{p1} = 10\lambda_{p2}$), and models without dimers, respectively. The figures on the left side from A to E are for the normalized period sensitivity of the Repressilator, and those on the right from F to J are for the Atkinson oscillator. The protein synthesis rate $\gamma = \alpha \cdot v/\lambda_m$ varies in the range 50–200 nM/min. The exact values of the parameters are set as follows: $\alpha = 50$–$200$ nM/min, $v = 0.2$ min$^{-1}$, and $\lambda_m = 0.2$ min$^{-1}$ for the Repressilator; and $\alpha = 5$–$20$ nM/min, $v = 2$ min$^{-1}$, and $\lambda_m = 0.2$ min$^{-1}$ for the Atkinson. The other parameters used are listed in Table 1.

## 4. Temperature Compensation

The normalized period sensitivity revealed that the variation of the kinetic parameters affects the period length of the oscillators. However, the periods of circadian rhythms are insensitive to thermal variations, even though many experiments have demonstrated that the reaction rates are highly dependent on the temperature [37]. This is the famous temperature compensation of circadian rhythms, but the mechanisms underlying this notable phenomenon are still unclear. Nevertheless, we know that the temperature sensitivity of the period of circadian clocks is closely related with two terms: the period and temperature sensitivities of the parameters [4].

### 4.1 Temperature Sensitivity of the Parameters and Temperature Coefficient

The temperature sensitivity of a parameter is described as the logarithmic change in the parameter with respect to the increase in ambient temperature [4]. If $s_j (j = 1, 2, \ldots, M)$ represents the temperature sensitivity of parameter $b_j$, then they are related by Eq. (9):

$$s_j = \frac{d \ln b_j}{dT} = \frac{1}{b_j} \cdot \frac{db_j}{dT}, \tag{9}$$

where $T$ represents the temperature. The temperature sensitivity of the parameters is always positive because the dynamic parameter $b_j$ increases with increasing temperature.

The value of the parameter $b_j$ is temperature-dependent, and the temperature coefficient $Q_{10}$ is commonly used to measure the thermal dependence of the reaction parameters. The temperature coefficient $Q_{10}$ is defined as the rate of change of the parameters when the temperature rises by 10°C; it can be written as

$$Q_{10} = \{b_j(T + \Delta T)/b_j(T)\}^{10/\Delta T}. \tag{10}$$

We can obtain the relationship between temperature coefficient $Q_{10}$ and temperature sensitivity of the parameters by:

$$\ln Q_{10} = 10 \cdot \frac{\ln b_j(T + \Delta T) - \ln b_j(T)}{\Delta T}$$

$$= 10 \cdot \frac{d \ln b_j}{dT} = 10 \cdot s_j. \tag{5.11}$$

Eq. (11) shows that the temperature sensitivity of a parameter can be expressed by the temperature coefficient $Q_{10}$ as $s_j = \ln Q_{10}/10$.

The experimental values of the temperature coefficient for the mRNA synthesis and degradation rates at 27°C and 17°C in *Arabidopsis* were recently reported, and the data followed a log-normal distribution [27]. We now show how to calculate a reasonable range for the temperature coefficient $Q_{10}$ of the mRNA reaction rate based on only the experimental data at 17°C and 27°C. Firstly, the experimental data for both the mRNA synthesis and degradation rate at 17°C and 27°C were mixed together to represent the sampling distribution of the temperature coefficient of the mRNA reaction rate at various temperatures. Then, we fitted the logarithmic values of the provided data with a normal distribution function to determine the feasible temperature coefficient region with high probability for the parameters of the genetic oscillators. In Fig. 3, which shows the original temperature coefficient data and fitting results, the red bars and the blue line represent the probability density distribution for the natural logarithm of the experimental temperature coefficient and the data fitting curve, respectively. The data fitting curve is a normal distribution with mean $\mu = 1.1171$ and standard deviation $\sigma = 0.4853$. The probability for values with a normal distribution falling within one standard deviation of the mean is about 68%, implying that the values of natural logarithmic $Q_{10}$ are mostly distributed in the interval $[\mu - \sigma = 0.6318, \mu + \sigma = 1.6024]$. The corresponding $Q_{10}$ was in the range $[e^{0.6318} = 1.8810, e^{1.6024} = 4.9649]$. Following this experimental data, the region of $Q_{10}$ for the transcriptional and degradation rates of mRNA in *Arabidopsis* was selected as [2.0, 5.0].

We cite the following references to verify that the range [2.0, 5.0] of the experimental data for *Arabidopsis* is also applicable for the mRNA transcription and degradation rates of bacteria. The half-life of the *cspA* transcript in *E. coli* is 20 s at most (extraordinarily unstable) at 37°C, but is about 30 min at 10°C (dramatically stabilized) [38], and the calculated temperature coefficient $Q_{10}$ for *cspA* mRNA is about 5.3. The *desA* transcript in the cyanobacteria *Synechococcus* sp. strainPCC 7002 is more stable at 22°C (half-life approximately 3.5-fold greater) than at 38°C, and the half-life of the *desB* gene is approximate 15-fold greater at 22°C than at 38°C [39], corresponding to $Q_{10}$ of 2.2 and 5.4, respectively. Typically, most biological reaction rates proceed with a temperature coefficient ($Q_{10}$) in the range [2.0, 3.0] [40]. The transcription of the *des* gene in *Bacillus subtilis* was improved by 10- to 15-fold when there was a shift of cultures from 37℃ to 20℃ (the range of the temperature coefficient $Q_{10}$ varies from 3.9 to 4.9) [41]. The decay rate of the *des* transcript as well as the *in vivo* degradation of *B. subtilis* bulk mRNA showed that stability increased about sixfold at 20℃ compared with 37℃ (a $Q_{10}$ of 3.0) [42-44]. Based on previous reports, most values of the temperature coefficient $Q_{10}$ of the mRNA reaction rate in bacteria also fall in the range [2.0, 5.0]. Thus, we assume that the temperature coefficient of mRNA in bacteria should not be very different from that in *Arabidopsis*.

At this point, we have decided the range of the temperature coefficient for the synthesis and degradation rates of mRNA, but whether this variation is also suitable for the temperature coefficient of the protein has yet to be determined. We summarize the experimental results provided in other studies in the following. Generally, the temperature coefficient $Q_{10}$ of proteins lies between 2.0 and 3.0; however, an experimental study showed that phosphatase has a value 5.0 for $Q_{10}$ and other processes controlled by the enzyme also displayed a much higher $Q_{10}$ [45], indicating a strong temperature dependence [46, 47]. The generation times of $S_1$ 55 protein in psychrotrophic bacterium *Arthrobacter* sp. were 19 h and 4 h 40 min at 10℃ and 20℃, respectively (a $Q_{10}$ of 4.1) [48]. LacI, a protein frequently used in bacterial networks, was degraded approximately 3−5 fold faster at 37℃ than at 25℃, corresponding temperature coefficient range from 2.5 to 3.8 [49]. The half-life of GmaR, protein-based thermosensors in *Listeria monocytogenes*, was determined to be at least 8 h at 30℃, and was reduced to 2−3 h at 37℃ in different situations [50]. The corresponding $Q_{10}$ value of GmaR varies from 4.1 to 7.2. Based on the experimental data provided in these references, a temperature coefficient for a protein within the range [2.0, 5.0] is also feasible.

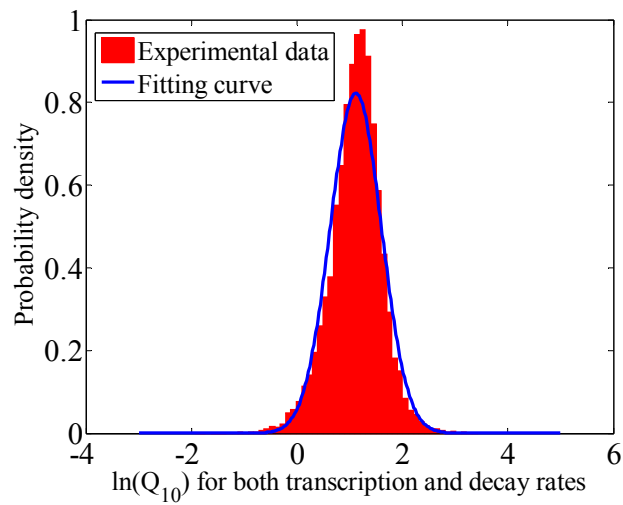

**Fig. 3.** Experimental data of the temperature coefficient and the related fitting curve. Distribution of the logarithmic experimental data for the temperature coefficient of the mRNA synthesis and degradation rates [27] and the fitting result. The probability density of the natural logarithm

of the temperature coefficient $Q_{10}$ is expressed by the bars, which clearly have a normal distribution, and the blue curve is the fitting result of the data with the normal distribution function. See the text for parameters of the fitted normal distribution.

### 4.2 Temperature Sensitivity of the Period

Here we consider the influence of different parameters on temperature compensation compared with previous research describing the realization of temperature compensation based on the Arrhenius equation [51, 52]. The previous research can be explained briefly as follows. The chemical rate equations, denoted by a set of reaction processes $R_j (j = 1, 2, ..., M)$, describe the time evolution of a physiological system. The Arrhenius equation can describe the influence of temperature on the rate parameter $b_j$ of an individual process $R_j$:

$$b_j = A_j e^{-\frac{E_j}{RT}}, \tag{5.12}$$

where $T$, $R$, $E_j$, and $A_j$ represent the temperature, gas constant, activation energy, and collision factor, respectively [51, 52]. The activation energy $E_j$ is a measure of how sensitively process $R_j$ responds to the temperature variation.

Temperature compensation requires that the following conditions must be satisfied [51, 52] based on the Arrhenius equation (Eq. (12)):

$$\frac{d \ln \tau}{dT} = \frac{1}{RT^2} \sum_{j=1}^{M} e_j E_j = 0, \tag{5.13}$$

where the normalized period sensitivity $e_j$ is also the so-called control coefficient [51, 52]. If the temperature compensation condition is satisfied, the control coefficient should be a set of positive and negative values since activation energies $E_j$ are positive. Eq. (13) shows that temperature compensation can be the result of a balancing between variations in the overall experimental activation energy.

Unlike in previous research [51, 52], we consider the influence of parametric temperature sensitivity ($s_j$) on temperature compensation ability in the feasible range indicated by experimental data after the normalized period sensitivity $e_j$ (also called the control coefficient) is calculated. We can calculate the normalized period sensitivity to the parameters by phase sensitivity analysis (Eq. (8)), and the range of the temperature sensitivity of the parameters can be determined according to the experimental data of the temperature coefficient. Accordingly, in our research, the period variation caused by parameter fluctuation is fixed, and we want to check whether temperature compensation can be obtained when the feasible temperature sensitivity of the parameters is selected. The normalized period sensitivity and the temperature sensitivity of the parameters determine the variation of the temperature sensitivity of the period, which is the key factor for explaining the mechanism of temperature compensation.

The period $\tau$ of the oscillator depends on the values of the various reaction rates $b_j (j = 1, 2, \ldots M)$ in the system, which in turn depend on the temperature $T$ of the environment. The temperature sensitivity of the period is defined as the logarithmic change in the period induced by a unit increase in the ambient temperature and can be written as [53]:

$$\frac{d \ln \tau}{dT} = \sum_{j=1}^{M} \frac{\partial \ln \tau}{\partial \ln b_j} \cdot \frac{d \ln b_j}{dT} = \sum_{j=1}^{M} e_j s_j, \tag{5.14}$$

where $e_j = \frac{\partial \ln \tau}{\partial \ln b_j}$ and $s_j = \frac{d \ln b_j}{dT}$ represent the normalized period sensitivity to the reaction rate parameter $b_j$ in Eq. (8) and the temperature sensitivity of the parameters in Eq. (9), respectively. Exact temperature compensation can be achieved when the rate of change of the period's temperature sensitivity is zero, that is, when Eq. (14) is equal to 0.

Thus we want to solve the following equation subject to the indicated constraints.

$$\min \left| \frac{d \ln \tau}{dT} \right| = \min \left| \sum_{j=1}^{M} e_j s_j \right|, \tag{5.15}$$

$$\text{subject to} \quad s_j = \ln Q_{10}/10; \; 2 \leq Q_{10} \leq 5.$$

Equation (15) can be solved with linear programming, and we can obtain the minimum temperature sensitivity of the period. High values indicate that the period of the oscillator is sensitive to temperature variation, whereas low values indicate that the period of the model is robust to temperature change. We compared the temperature compensation ability for different mathematical models of both the Repressilator and the Atkinson oscillator when the protein synthesis rate at full activation ($\gamma = \alpha v/\lambda_m$) was between 50 and 200 nM/min. The parameters are identical to those in the calculation of normalized period sensitivity shown in Fig. 2. Figure 4 presents calculation results for different models of genetic oscillators, showing how the protein dimerization and cooperative stability (nonlinear protein degradation) affect the temperature sensitivity of the period with realizable *in vivo* data for the Repressilator and the Atkinson oscillator. It can be seen that the models without dimeric proteins (dash-dotted black lines) for both the Repressilator and the Atkinson oscillator have the worst temperature compensation ability in most situations. The solid-blue lines in Fig. 4 represent the temperature compensation ability of the Repressilator and the Atkinson oscillator with linear protein degradation observed for all proteins. The dashed-red ($\lambda_{p1} = 5\lambda_{p2}$) and dotted-green ($\lambda_{p1} = 10\lambda_{p2}$) lines show the oscillators' cooperative stability in one or several proteins. Figures 4A, B, and C illustrate that nonlinear protein degradation can happen in all three proteins (Eq. (2)), two proteins (Eq. (A1)), and only one protein (Eq. (A2)) of the Repressilator network, respectively. The results for the Atkinson oscillator with two proteins (Eq. (4)), protein 2 (Eq. (A3)) and protein 1 (Eq. (A4)) having different degrees of cooperativity are shown in Figs. 4D, E, and F, respectively. The theoretical results for the minimum temperature sensitivity of the period show that cooperative stability indeed improves temperature compensation ability. The temperature compensation of the oscillator improves as the degree of cooperativity becomes larger and more proteins have cooperative stability. We also provide data on the minimum $|\frac{d \ln \tau}{dT}|$ for different models of the two oscillators when the translation rate $v$ and mRNA degradation rate $\lambda_m$ vary, showing results that are similar to those in Fig. 4 (see Supplemental Material). The theoretical analysis and calculation show that in most situations, protein dimerization can improve the temperature compensation ability. Furthermore, compared with the linear protein degradation model, the nonlinear protein degradation model is more appropriate to describe the mechanism of temperature compensation for the two oscillators.

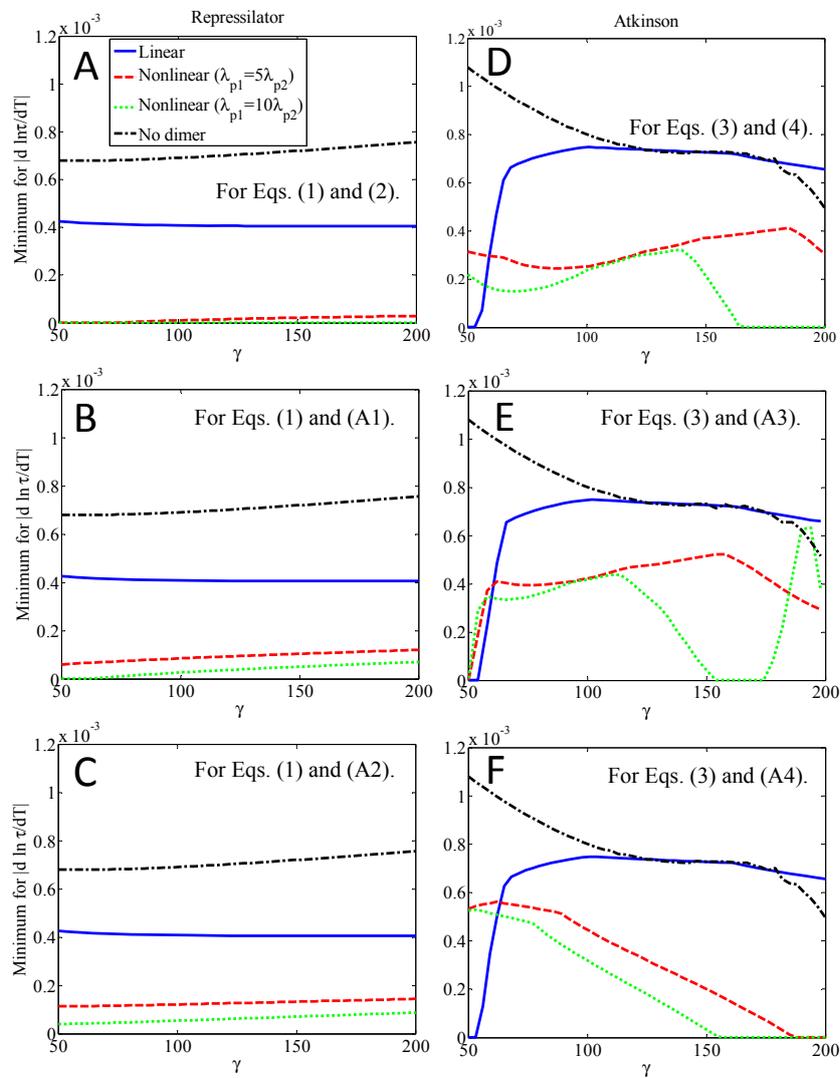

**Fig. 4.** The minimum of $\left|\frac{d\ln\tau}{dT}\right|$ for different models of both the Repressilator and the Atkinson oscillator when $\alpha$ varies. The left figures show the Repressilator results, and the right figures show the temperature compensation ability for different models of the Atkinson oscillator. (A) and (D): All proteins in the oscillators can degrade linearly or nonlinearly (expressed by Eqs. (2) and (4), respectively). (B): Two proteins of the Repressilator can degrade linearly or nonlinearly, and the remaining protein's degradation is linear (shown by Eq. (A1)). (C) Only one protein of the Repressilator shows cooperative stability, and the others undergo linear degradation (described by Eq. (A2)). (E): Protein 2 in the Atkinson oscillator can undergo linear or nonlinear degradation, protein 1 undergoes linear degradation (Eq. (A3)). (F) Opposite to the situation of (E) (results of Eq. (A4)). The solid (blue), dashed (red), and dotted (green) lines represent the temperature compensation ability for all or some proteins undergoing linear degradation ($\lambda_{p1} = \lambda_{p2}$), or nonlinear protein degradation with $\lambda_{p1} = 5\lambda_{p2}$, and $\lambda_{p1} = 10\lambda_{p2}$, respectively. The minimum $\left|\frac{d\ln\tau}{dT}\right|$ of the mathematical model without protein dimerization expressed by Eqs. (1) (Repressilator) and (3) (Atkinson) is shown as dash-dot (black) line. The exact parameter values are as follows: $\alpha = 50 - 200$ nM/min, $v = 0.2$ min$^{-1}$, and $\lambda_m = 0.2$ min$^{-1}$ for the Repressilator; $\alpha = 5 - 20$ nM/min, $v = 2$ min$^{-1}$, and $\lambda_m = 0.2$ min$^{-1}$ for the Atkinson oscillator. The values of other parameters are listed in Table 1. The corresponding protein synthesis rate is $\gamma = \alpha \cdot v/\lambda_m = 50 - 200$ nM/min.

To study the temperature compensation ability, we applied realizable *in vivo* data to the Repressilator and the Atkinson oscillator to analyze how dimeric proteins and cooperative stability affect the temperature compensation on the basis of period sensitivity analysis. The results show that the mathematical models (Eqs. (1) and (3)) without dimeric proteins have the worst temperature compensation. The cooperative stability mechanism also makes temperature compensation much better than models with linear protein degradation. The more proteins that show cooperative stability and the higher the degree of protein cooperativity, the easier it is for the oscillator to achieve temperature compensation.

## 5. Discussion

Circadian clocks exhibit temperature compensation. Therefore, the temperature sensitivity of their period (Eq. (14)) should be close to zero. In an effort to explain the mechanism underlying temperature compensation, we analyzed the effects of protein dimerization and cooperative stability on the temperature compensation ability of two oscillators. The period's temperature sensitivity depends on the normalized period sensitivity and temperature sensitivity of the parameters. We calculated the period sensitivity to the parameters by performing phase sensitivity analysis and we determined the feasible range of the temperature coefficient in accordance with recent biological experiments [27]. Given the values and the constraints, we then computed the attainable minimum for the temperature sensitivity of the period using the linear programming method. The theoretical results (Figs. 4, S9, and S10) show that it is more difficult to obtain temperature compensation without dimers than with dimeric proteins. From the same results, we can conclude that nonlinear protein degradation indeed improves the temperature compensation ability of oscillators compared with linear protein degradation, in most situations. Figures 4, S9, and S10 also show that the temperature compensation improves as the cooperativity between the dimers and monomers increases and more proteins exhibit cooperative stability.

In addition to our investigation of the Repressilator and the Atkinson oscillator, we also derived theoretical results for the Goodwin oscillator, a well-studied model relevant to circadian oscillations [44, 54-57], to show the generality of our findings. Figure 5 shows the simulation results for the Goodwin oscillator expressed by Eq. (A5), in which two proteins exist as both dimers and monomers. In Fig. 5, the linear protein degradation ($\lambda_{p1} = \lambda_{p2}$), the nonlinear protein degradation with $\lambda_{p1} = 5\lambda_{p2}$, and $\lambda_{p1} = 10\lambda_{p2}$ are shown as the solid (blue), dashed (red), and dotted (green) lines, respectively. Compared with linear protein degradation (solid-blue line), the models with nonlinear protein degradation exhibit better temperature compensation. If the degree of cooperativity is larger ($\lambda_{p1} = 10\lambda_{p2}$ vs. $\lambda_{p1} = 5\lambda_{p2}$), the Goodwin oscillator exhibits better temperature compensation ability (dotted-green line vs. dashed-red line). In the supplemental material, we also provide the calculation result for the Goodwin oscillator expressed in Eq. (A5) when $v$ and $\lambda_m$ vary, and the results are very similar to Fig. 5.

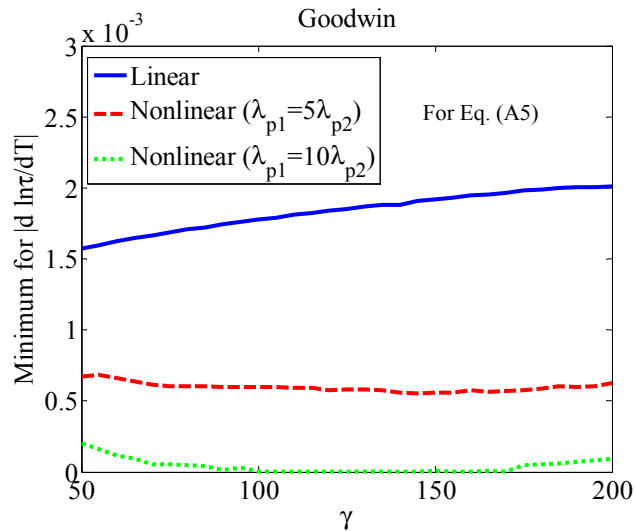

**Fig. 5**. The minimum of $|\frac{d \ln \tau}{d T}|$ for the Goodwin oscillator expressed by Eq. (A5) when $\alpha$ varies. The solid (blue), dashed (red), and dotted (green) lines represent the temperature compensation ability for linear protein degradation $\lambda_{p1} = \lambda_{p2}$, and in nonlinear protein degradation with $\lambda_{p1} = 5\lambda_{p2}$, and $\lambda_{p1} = 10\lambda_{p2}$, respectively. The exact parameters in Eq. (A5) are: $\alpha = 5 - 20$ nM/min, $v = 2$ min$^{-1}$, $\lambda_m = 0.2$ min$^{-1}$, $\lambda_{p1} = 0.2$ min$^{-1}$, $l = 1000$, $k = 3$ nM, $n = 10$, $K_d = 10$ nM, $k_5 = 1$ min$^{-1}$. The corresponding protein synthesis rate is $\gamma = \alpha v/\lambda_m = 50 - 200$ nM/min.

To elucidate why the models with cooperative stability can obtain better temperature compensation, we need to consider the effect of the normalized period sensitivity, shown in Fig. 2. Without dimeric proteins in the models, the period sensitivities to the mRNA degradation rate (Figs. 2C and H) and the monomeric degradation rate (Figs. 2D and I) are much higher. However, the differences between the models' normalized period sensitivity to the transcription rate $\alpha$ and the translation rate $v$ are relatively small. Therefore, the temperature compensation ability for the no-dimer models is weaker than that for the models with dimeric proteins (Fig. 4). The same explanation is also applicable to the results for the same models shown in Figs. S9 and S10, which are provided in the supplemental material. The normalized period sensitivities vary considerably between linear and nonlinear protein degradation models in the Repressilator, resulting in different temperature compensation abilities. The Repressilator's normalized period sensitivity to $\alpha$ (Fig. 2A), $v$ (Fig. 2B), $\lambda_m$ (Fig. 2C), and $\lambda_{p2}$ (Fig. 2E) in the nonlinear model is more robust to perturbations, meaning closer to zero. Figure 2D shows that large differences were not observed between the period variations of the linear and nonlinear models caused by disturbances to $\lambda_{p1}$. Thus, the Repressilator with nonlinear protein degradation can much more easily achieve temperature compensation. As for the Atkinson oscillator, when the protein synthesis rate is small, the reason for more robust temperature compensation in the nonlinear model is almost the same as that for the Repressilator. In the range of large protein synthesis rates, the normalized period sensitivities to transcription and translation rates ($\alpha$ and $v$) (Figs. 2F and G) in the Atkinson oscillator's linear model are close to zero and slightly greater than zero, respectively. Figures 2H and J show that the normalized period sensitivities of the Atkinson oscillator with the linear model for parameters $\lambda_m$ and $\lambda_{p2}$ are much less than zero. In Fig. 2I, the period of the linear model will decrease slightly when there are perturbations to $\lambda_{p1}$, meaning that the period sensitivity is only slightly less than zero. Only the period sensitivity to $v$ (Fig. 2G) is positive but the value is very small; therefore, it is difficult for the Atkinson oscillator with a linear model to achieve temperature compensation. The nonlinear protein degradation model for the Atkinson oscillator changes the direction of the period variations. For example, in the range of the large protein synthesis rate, perturbations to parameter $\alpha$ (Fig. 2F) will increase the period of the linear protein degradation model, while decreasing the period of the nonlinear model. The nonlinear model's normalized period sensitivities for $\alpha$ and $v$ (Figs. 2F and G) are much less than zero; while the normalized period sensitivities for parameters $\lambda_m$ and $\lambda_{p1}$ (Figs. 2H and I) are greater than zero. Although the absolute values of the normalized period sensitivities for parameters $\alpha$, $v$, $\lambda_m$, and $\lambda_{p1}$ in the nonlinear protein degradation model are greater than those in the linear model, the existence of negative and positive values makes it easier to reduce the variation of the period on the whole. The absolute value of the nonlinear model's period sensitivity to the dimeric protein degradation rate $\lambda_{p2}$ is much smaller than that of the linear model. Therefore, it is easier for the Atkinson oscillator with cooperative stability (nonlinear protein degradation) to achieve temperature compensation.

Regardless of the reason, for the Repressilator and the Atkinson oscillator, the presence of dimeric proteins and cooperative stability in protein degradation process greatly change the period sensitivity to the parameters in different models. Mathematical models without protein dimerization do not have a dimeric protein degradation rate $\lambda_{p2}$ (Figs. 2E and J), and the normalized period sensitivities to $\lambda_{p2}$ are very different in the linear and nonlinear protein degradation models. In view of the theoretical analysis presented in Figs. 2 and S1–S8, we can conclude that cooperative stability affects the period sensitivity to $\lambda_{p2}$, which in turn plays an important role in the temperature compensation ability of these oscillators. In other words, the cooperative stability incorporated in protein degradation confers better temperature compensation on the basis of biologically feasible parameters. It is expected that this mechanism will be implemented *in vivo* because it is a prevalent mechanism in cells. If it is necessary to design synthetic genetic oscillators with low period sensitivity to temperature, nonlinear protein degradation for the circuits should be considered.

## Appendix A

The ODEs for the Repressilator when protein 2 and protein 3 have cooperative stability, and the monomeric and dimeric protein degradation of the left protein is linear:

$$\frac{dm^{(1)}}{dt} = \alpha g_r(p_2^{(3)}) - \lambda_m m^{(1)},$$

$$\frac{dp^{(1)}}{dt} = vm^{(1)} - \lambda_{p1} p^{(1)},$$

$$\frac{dm^{(2)}}{dt} = \alpha g_r(p_2^{(1)}) - \lambda_m m^{(2)},$$

$$\frac{dp^{(2)}}{dt} = vm^{(2)} - (\lambda_{p1} p_1^{(2)} + 2\lambda_{p2} p_2^{(2)}),$$

$$\frac{dm^{(3)}}{dt} = \alpha g_r(p_2^{(2)}) - \lambda_m m^{(3)},$$

$$\frac{dp^{(3)}}{dt} = vm^{(3)} - (\lambda_{p1} p_1^{(3)} + 2\lambda_{p2} p_2^{(3)})$$

(A1)

The ODEs for the Repressilator when only protein 3 shows cooperative stability and the monomeric and dimeric protein degradation of the other two proteins are linear:

$$\frac{dm^{(1)}}{dt} = \alpha g_r(p_2^{(3)}) - \lambda_m m^{(1)},$$

$$\frac{dp^{(1)}}{dt} = vm^{(1)} - \lambda_{p1} p^{(1)},$$

(A2)

$$\frac{dm^{(2)}}{dt} = \alpha g_r(p_2^{(1)}) - \lambda_m m^{(2)},$$

$$\frac{dp^{(2)}}{dt} = vm^{(2)} - \lambda_{p1} p^{(2)},$$

$$\frac{dm^{(3)}}{dt} = \alpha g_r(p_2^{(2)}) - \lambda_m m^{(3)},$$

$$\frac{dp^{(3)}}{dt} = vm^{(3)} - \left(\lambda_{p1} p_1^{(3)} + 2\lambda_{p2} p_2^{(3)}\right)$$

The ODEs for the Atkinson oscillator when protein 2 shows cooperative stability, and the monomeric and dimeric protein degradation of protein 1 is linear:

$$\frac{dm^{(1)}}{dt} = \alpha g_r(p_2^{(2)}) g_a(p_2^{(1)}) - \lambda_m m^{(1)},$$

$$\frac{dm^{(2)}}{dt} = \alpha g_a(p_2^{(1)}) - \lambda_m m^{(2)},$$

$$\frac{dp^{(1)}}{dt} = vm^{(1)} - \lambda_{p1} p^{(1)},$$

$$\frac{dp^{(2)}}{dt} = vm^{(2)} - \left(\lambda_{p1} p_1^{(2)} + 2\lambda_{p2} p_2^{(2)}\right),$$

(A3)

The ODEs for the Atkinson oscillator when protein 1 shows cooperative stability, and the monomeric and dimeric protein degradation of protein 2 is linear:

$$\frac{dm^{(1)}}{dt} = \alpha g_r(p_2^{(2)}) g_a(p_2^{(1)}) - \lambda_m m^{(1)},$$

$$\frac{dm^{(2)}}{dt} = \alpha g_a(p_2^{(1)}) - \lambda_m m^{(2)},$$

$$\frac{dp^{(1)}}{dt} = vm^{(1)} - \left(\lambda_{p1} p_1^{(1)} + 2\lambda_{p2} p_2^{(1)}\right),$$

$$\frac{dp^{(2)}}{dt} = vm^{(2)} - \lambda_{p1} p^{(2)},$$

(A4)

The ODEs for the Goodwin oscillator with cooperative stability for each protein (refer to [58] for details of the model):

$$\frac{dm}{dt} = \alpha g_r(z_2) - \lambda_m m,$$

$$\frac{dp}{dt} = vm - (\lambda_{p1} p_1 + 2\lambda_{p2} p_2),$$

$$\frac{dz}{dt} = k_5 p_2 - (\lambda_{p1} z_1 + 2\lambda_{p2} z_2),$$

(A5)

where the variables $m$, $p$, and $z$ can be interpreted as the total concentrations of mRNA, the corresponding protein, and a transcriptional inhibitor, respectively. The subscript 1 indicates monomers, and 2 indicates dimers. The relationships between the concentrations of monomers and dimers and the total protein concentrations are $p = p_1 + 2p_2$ and $z = z_1 + 2z_2$. $k_5$ is the synthesis rate of protein $z$. The other parameters in the Goodwin oscillator have the same meaning with those in the models of the Repressilator and Atkinson.

## Appendix B

The relationship between the position of the point $\boldsymbol{y}^\varsigma$ in Eq. (5) on the trajectory and the phase $\psi$ of the oscillator should have one-to-one mapping in one period. The phase $\psi$ is calculated by using the elapsed time $t$ from the reference point (zero phase) to the current point modulo the period $\tau$ of the oscillator. Because the increase of the phase $\psi$ is constant, we can use a differential equation to define the phase as an evolution process through time, as follows:

$$\frac{d\psi(\boldsymbol{y}^\varsigma(t,\boldsymbol{b}))}{dt} = 1, \quad \psi(y_0^\varsigma) = 0, \tag{B1}$$

where the phase of the reference point $y_0^\varsigma$ on the orbit is 0. Using the phase reduction method, we can represent the dynamical systems of the genetic oscillators by Eq. (B1) with respect to the phase.

If the trajectory of the oscillator is interrupted, the phase will deviate from the original orbit. Therefore, we first considered that the states, that is, the concentration of the protein or mRNA, of the oscillators are disrupted by infinitesimal perturbations. As time progresses, the perturbed point along the orbit finally approaches a different position (phase) of the nominal limit cycle, incurring a phase shift between two positions on the limit cycle. The set of phase shifts induced by small short-lived stimuli at different times (phases) of the orbit is PRC, and the symbol $\boldsymbol{U}$ represents a vector of the PRCs caused by impulse perturbations to all the states of the oscillators. Although the PRC had the simplest phase analysis, it is necessary for studying more complex phase sensitivities. We assumed that one solution $y_i^\varsigma(t)$ ($i$=1, 2, ..., $N$) on the nominal system was disrupted by infinitesimal stimuli, which causes a small phase difference $\Delta\psi$. The following equation can explain the mathematical meaning of the PRC $U_i$:

$$U_i(\boldsymbol{y}^\varsigma(t)) = \frac{\partial \psi}{\partial y_i^\varsigma}(\boldsymbol{y}^\varsigma(t)). \tag{B2}$$

There exist several approaches, such as finite difference, adjoint Green functions, and Malkin's methods, to calculate the state sensitivity of the dynamic model expressed in Eq. (B2). Here, we adopted Malkin's approach [59] to compute the PRCs $\boldsymbol{U}$ according to

$$\frac{d}{dt}U(y^c(t)) = -J^T(y^c(t)) \cdot U(y^c(t)),$$

$$s.t. \ U(y^c(0)) \cdot f(y^c(0)) = 1,$$

(B3)

where $J^T(y^c(t))$ represents the transposition of the Jacobian matrix $J(y^c(t))$ ($J \in R^{N \times N}$, $J_{ij} = \partial f_i / \partial y_j$), and $U(y^c(0)) \cdot f(y^c(0)) = 1$ is the initial boundary condition of the differential equation. This differential equation must be integrated backwards from the final time to the initial time.

## Acknowledgements

This work was supported by a Grant-in-Aid for Scientific Research on Innovative Areas "Synthetic biology for the comprehension of biomolecular networks".